\documentclass[10pt,journal]{IEEEtran}
\usepackage[latin9]{inputenc}
\usepackage{amsmath}
\usepackage{amssymb}
\usepackage{graphicx}

\makeatletter

\providecommand{\tabularnewline}{\\}

\makeatother

\begin{document}

\title{Bit-Level Soft-Decision Decoding of Triple-Parity Reed-Solomon Codes
through Automorphism Groups}

\author{Vo Tam Van, Seiichi Mita,~Jing (Tiffany) Li,~Chau Yuen, ~Yong
Liang Guan }
\maketitle
\begin{abstract}
This paper discusses bit-level soft decoding of triple-parity Reed-Solomon
(RS) codes through automorphism permutation. A new method for identifying
the automorphism groups of RS binary images is first developed. The
new algorithm runs effectively, and can handle more RS codes and capture
more automorphism groups than the existing ones. Utilizing the automorphism
results, a new bit-level soft-decision decoding algorithm is subsequently
developed for general $(n,n-3,4)$ RS codes. Simulation on $(31,28,4)$
RS codes demonstrates an impressive gain of more than 1 dB at the
bit error rate of $10^{-5}$ over the existing algorithms. \end{abstract}
\IEEEpeerreviewmaketitle

\section{Introduction}

Reed-Solomon (RS) codes, with their renowned Berlekamp-Massey and
Forney algorithms, boast robust error correcting capability against
bit-flip errors such as large amplitude fluctuation errors and burst
errors. However, to fully harness their power on additive white Gaussian
noise (AWGN) channels requires effective soft decoding, which has
been a research focus in recent years (see, for example, \cite{Narayanan,Vardy}).

Some algorithms, such as sorting- and scheduling- based message-passing
algorithms, are designed for general RS codes \cite{Narayanan,Vardy}.
Not restricting the underlying code specifications, these algorithms
achieve generality at the cost of a rather high level of computational
complexity. In comparison, the algorithms that target specific classes
of RS codes can explore specific code structures to effectively reduce
the complexity and/or improve the performance. Among the latter type,
a notable example is the bit-level automorphism-based decoder developed
in \cite{FLim}, which is designed for double-parity RS codes. Working
especially well for short-length (double-parity) RS codes, the algorithm
in \cite{FLim} is shown to be capable of a performance within $0.3$dB
to the maximum likelihood decoder (MLD) at a bit error rate (BER)
of $10^{-5}$. Despite its efficiency, however, this algorithm cannot
handle RS codes of more than two parity symbols, in part due to its
incapability of obtaining their automorphism groups.

This paper studies bit-level soft decoding of triple-parity RS codes.
Motivated by the algorithm in \cite{FLim}, we first investigate the
automorphism groups of the binary images of a general $(n,n-3,4)$
RS code, and propose a heuristic search permutation algorithm to identify
the automorphism groups. Through concrete examples, we show that our
algorithm works effectively, can handle codes that previous algorithms
cannot, and find automorphism groups that previous algorithms cannot.
Exploiting these automorphism groups and integrating them with the
general-purpose soft decoder (rather than the hard decoder in \cite{FLim}),
we further develop a new bit-level soft decoding algorithm, termed
\emph{permutation sum-product algorithm} (PSPA), for triple-parity
RS codes. Simulation demonstrates that our algorithm can noticeably
outperform the existing algorithms, such as conventional hard-decision
decoding (HDD) \cite{Berlekamp} and the soft-decision SPA \cite{Pearl}.
The proposed new algorithm is most efficient for short-length RS codes,
and can be extended to soft decode concatenated or compound codes
that use short-length triple-parity RS codes as the component code.

The rest of the paper is organized as follows. Section II introduces
the binary image of RS codes. Section III discusses automorphism groups
and details our searching algorithms for $(n,n-3,4)$ RS binary images.
Section IV presents a new soft decoding method that effectively combines
automorphism permutations and soft-decision sum product algorithm
(SPA). Section V demonstrates simulation results, and Section VI concludes
the paper.

\section{Binary Images of RS Codes}

Let $\gamma=[\gamma_{1},\gamma_{2},\cdots,\gamma_{m}]$ be a basis
of $\mathbb{F}_{2^{m}}$ over $\mathbb{F}_{2}$. Let the code length
be $n=2^{m}-1$. The binary image of a double-parity $(n,n-2,3)$
RS codeword $\mathbf{c}=[c_{0},c_{1},\cdots,c_{n-1}]$ is an $m\times n$
binary matrix: 
\begin{equation}
B_{M}(\mathbf{c}):=\left[\begin{array}{ccccc}
c_{1,0} & c_{1,1} & c_{1,2} & \cdots & c_{1,n-1}\\
c_{2,0} & c_{2,1} & c_{2,2} & \cdots & c_{2,n-1}\\
\vdots & \vdots & \vdots & \vdots & \vdots\\
c_{m,0} & c_{m,1} & c_{m,2} & \cdots & c_{m,n-1}
\end{array}\right],\label{eq:binaryImage}
\end{equation}
where $c_{j}=c_{1,j}\gamma_{1}+c_{2,j}\gamma_{2}+\cdots+c_{m,j}\gamma_{m}$
and $c^{(i)}=[c_{i,0},c_{i,1},c_{i,2},\cdots,c_{i,n-1}]$ for all
$j\in{\normalcolor \mathbb{Z}}_{n}$ and $1\leq i\leq m$.

The parity check matrix of the $(n,n-2,3)$ RS binary image with zeros
$\{1,\alpha\}$ can be presented by a $2m\times m$ polynomial in
the ring $\mathbb{F}_{2}[x]/(x^{n}-1)$ \cite{FLim}: 
\begin{equation}
\left[\begin{array}{cccc}
\theta_{1}(x) & 0 & \cdots & 0\\
0 & \theta_{1}(x) & \cdots & 0\\
\vdots & \vdots & \vdots & \vdots\\
0 & 0 & \cdots & \theta_{1}(x)\\
\theta_{\varepsilon}(x)x^{u_{1}} & \theta_{\varepsilon}(x)x^{u_{2}} & \cdots & \theta_{\varepsilon}(x)x^{u_{m}}\\
\vdots & \vdots & \vdots & \vdots\\
\theta_{\varepsilon}(x)x^{u_{1}+m-1} & \theta_{\varepsilon}(x)x^{u_{2}+m-1} & \cdots & \theta_{\varepsilon}(x)x^{u_{m}+m-1}
\end{array}\right],\label{eq:parityBinaryImage}
\end{equation}
where $\varepsilon=\alpha^{-1}$, $\theta_{\varepsilon}(x)$ is known
as the \textit{idempotent} \cite{MacWilliams}, and $\theta_{1}(x)=1+x+x^{2}+\cdots+x^{n-1}$.
Examples of vectors $\mathbf{u}=(u_{1},u_{2},\cdots,u_{m})\in\mathbb{Z}_{n}^{m}$
are listed in Table \ref{table:vectorU-1}.

\begin{table}[htbf]
\centering \caption{$\mathbf{u}$ vectors computed for $\gamma=[\gamma_{1},\gamma_{2},\cdots,\gamma_{m}]$.}

\vspace{-0.2cm}

\label{table:vectorU-1} %
\begin{tabular}{|l|c|c|}
\hline 
 & Vector $\mathbf{u}$  & Primitive element $\alpha$\tabularnewline
\hline 
$\mathbb{F}_{2^{3}}$  & $[2,1,0]^{\textrm{T}}$  & $\alpha^{3}=\alpha+1$\tabularnewline
\hline 
$\mathbb{F}_{2^{4}}$  & $[2,1,0,14]^{\textrm{T}}$  & $\alpha^{4}=\alpha+1$\tabularnewline
\hline 
$\mathbb{F}_{2^{5}}$  & $[30,29,28,27,26]^{\textrm{T}}$  & $\alpha^{5}=\alpha^{2}+1$\tabularnewline
\hline 
$\mathbb{F}_{2^{6}}$  & $[4,3,2,1,0,62]^{\textrm{T}}$  & $\alpha^{6}=\alpha+1$\tabularnewline
\hline 
\end{tabular}
\end{table}

\section{Automorphisms of Triple-Parity RS Codes\label{sec:Aut3parity}}

Previous work has reported all the automorphism groups for the $(7,4,4)$
RS code over $\mathbb{F}_{2^{3}}$ and part of the automorphism groups
for the $(15,12,4)$ RS code over $\mathbb{F}_{2^{4}}$ \cite{FLim}.
Here, we focus on general $(n,n-3,4)$ RS codes with zeros $\left\{ 1,\alpha,\alpha^{2}\right\} $
over $\mathbb{F}_{2^{m}}$ for all $m\geq4$. Since, $\alpha$ and
$\alpha^{2}$ belong to the same cyclotomic coset, the parity check
matrix of $(n,n-3,4)$ RS codes is represented by the following $3m\times m$
polynomial matrix in the ring $\mathbb{F}_{2}[x]/(x^{n}-1)$.

\begin{equation}
\left[\begin{array}{cccc}
\theta_{1}(x) & 0 & \cdots & 0\\
0 & \theta_{1}(x) & \cdots & 0\\
\vdots & \vdots & \vdots & \vdots\\
0 & 0 & \cdots & \theta_{1}(x)\\
\theta(x)x^{u_{1}^{(1)}} & \theta(x)x^{u_{2}^{(1)}} & \cdots & \theta(x)x^{u_{m}^{(1)}}\\
\vdots & \vdots & \vdots & \vdots\\
\theta(x)x^{u_{1}^{(1)}+m-1} & \theta(x)x^{u_{2}^{(1)}+m-1} & \cdots & \theta(x)x^{u_{m}^{(1)}+m-1}\\
\theta(x)x^{u_{1}^{(2)}} & \theta(x)x^{u_{2}^{(2)}} & \cdots & \theta(x)x^{u_{m}^{(2)}}\\
\vdots & \vdots & \vdots & \vdots\\
\theta(x)x^{u_{1}^{(2)}+m-1} & \theta(x)x^{u_{2}^{(2)}+m-1} & \cdots & \theta(x)x^{u_{m}^{(2)}+m-1}
\end{array}\right],\label{eq:parityBinaryImage3parity}
\end{equation}
where $\theta(x)$ is short for $\theta{}_{\varepsilon}(x)$. The
vectors $\mathbf{u}^{(1)}=[u_{1}^{(1)},u_{2}^{(1)},\cdots,u_{m}^{(1)}]$
and $\mathbf{u}^{(2)}=[u_{1}^{(2)},u_{2}^{(2)},\cdots,u_{m}^{(2)}]$$\in\mathbb{F}_{n}^{m}$
can be computed using the method introduced in \cite{FLim}, and the
results are listed in Table \ref{table:vectorU1-U2}.

\begin{table}[htbf]
\centering \caption{Common vectors $\mathbf{u}^{(1)}$ and $\mathbf{u}^{(2)}$ for $(n,n-3,4)$
RS codes.}

\vspace{-0.2cm}

\label{table:vectorU1-U2} %
\begin{tabular}{|l|c|c|}
\hline 
 & Vector $\mathbf{u}^{(1)}$  & Vector $\mathbf{u}^{(2)}$\tabularnewline
\hline 
$\mathbb{F}_{2^{3}}$  & $[2,1,0]^{\textrm{T}}$  & $[2,5,1]^{\textrm{T}}$\tabularnewline
\hline 
$\mathbb{F}_{2^{4}}$  & $[2,1,0,14]^{\textrm{T}}$  & $[2,9,1,8]^{\textrm{T}}$\tabularnewline
\hline 
$\mathbb{F}_{2^{5}}$  & $[30,29,28,27,26]^{\textrm{T}}$  & $[30,14,29,13,28]^{\textrm{T}}$\tabularnewline
\hline 
$\mathbb{F}_{2^{6}}$  & $[4,3,2,1,0,62]^{\textrm{T}}$  & $[4,35,3,34,2,33]^{\textrm{T}}$\tabularnewline
\hline 
$\mathbb{F}_{2^{7}}$  & $[6,5,4,3,2,1,0]^{\textrm{T}}$  & $[6,69,5,68,4,67,3]^{\textrm{T}}$\tabularnewline
\hline 
$\mathbb{F}_{2^{8}}$  & $[4,3,2,1,0,254,253,252]^{\textrm{T}}$  & $[4,131,3,130,2,129,1,128]^{\textrm{T}}$\tabularnewline
\hline 
$\mathbb{F}_{2^{9}}$  & $\begin{array}{c}
[510,509,508,507,506,\\
505,504,503,502]^{\mathrm{T}}
\end{array}$  & $\begin{array}{c}
[510,254,509,253,508,\\
252,507,251,506]^{\mathrm{T}}
\end{array}$\tabularnewline
\hline 
$\mathbb{F}_{2^{10}}$  & $\begin{array}{c}
[6,5,4,3,2,1,0,\\
1022,1021,1020]^{\mathrm{T}}
\end{array}$  & $\begin{array}{c}
[6,517,5,516,4,515,\\
3,514,2,513]^{\mathrm{T}}
\end{array}$\tabularnewline
\hline 
\end{tabular}
\end{table}

Define the symmetric group on the set $\left\{ 1,2,\cdots,m\right\} $
as $\Omega{}_{m}$. A permutation $\sigma\in\Omega_{m}$ is represented
using the cycle notation (e.g., the permutation $\sigma=(1,2,4)$
means $\sigma(1)=2$, $\sigma(2)=4$, and $\sigma(4)=1$). Let \emph{id}
denote the identity permutation on $m$ indices $\left\{ 1,2,\cdots,m\right\} $.
Let $a_{1},a_{2},\cdots,a_{m}\in\mathbb{Z}_{n}$ and $l\in\mathbb{Z}_{m}$.
The parameter $(\sigma,(a_{1},a_{2},\cdots,a_{m}),l)$ completely
describes the permutation $\rho$, such that $[i,j]\mapsto[\sigma(i),j2^{l}+a_{i}]$.

We present the basic idea of identifying the automorphism through
the following example.

\vspace{0.5cm}
 \textbf{Example 1: {[}$(15,12,4)$ RS code{]}} Let $\alpha$ be a
fixed primitive element satisfying $\alpha^{4}=\alpha+1$. Consider
the $(15,12,4)$ RS binary image with zeros $\left\{ 1,\alpha,\alpha^{2}\right\} $,
imaged under the canonical basis.

Consider a specific permutation $\rho$ that is described by the parameters
$\sigma=(1,2)(3,4)$, $(a_{1},a_{2},a_{3},a_{4})$ and $l$. It can
be verified that the permutations $(\sigma,(a_{1},a_{2},\cdots,a_{m}),l)$
and $(\sigma,(a_{1},a_{2},\cdots,a_{m})+a,l)$ belong to the same
automorphism group. Therefore, it is sufficient to fix parameter $a_{1}=0$,
then $\rho$ maps the indices 
\begin{equation}
\begin{array}{cc}
[1,j]\mapsto[2,j2^{l}], & [2,j]\mapsto[1,j2^{l}+a_{2}]\\
{}[3,j]\mapsto[4,j2^{l}+a_{3}], & [4,j]\mapsto[3,j2^{l}+a_{4}]
\end{array}\label{eq:RS15bin2}
\end{equation}
The parity check matrix of the $(15,12,4)$ RS binary image is invariant
under the permutation specified in (\ref{eq:RS15bin2}), if and only
if the following generator matrix in the ring $\mathbb{F}_{2}[x]/(x^{15}-1)$
is invariant under permutation specified in (\ref{eq:RS15bin2}):
\begin{equation}
\left[\begin{array}{cccc}
0 & \theta(x)x^{b_{12}} & \theta(x)x^{b_{13}} & \theta(x)x^{b_{14}}\\
\theta(x)x^{b_{21}} & 0 & \theta(x)x^{b_{23}} & \theta(x)x^{b_{24}}\\
\theta(x)x^{b_{31}} & \theta(x)x^{b_{32}} & 0 & \theta(x)x^{b_{34}}\\
\theta(x)x^{b_{41}} & \theta(x)x^{b_{42}} & \theta(x)x^{b_{43}} & 0
\end{array}\right],\label{eq:RS15polymatrix}
\end{equation}
where $(b_{12},b_{13},b_{14})=(7,12,1),$ $(b_{21},b_{23},b_{24})=(8,14,4),$
$(b_{31},b_{32},b_{34})=(14,0,6),$ and $(b_{41},b_{42},b_{43})=(4,6,7)$.

Note that permuting $\theta(x)x^{r}$ by $j\mapsto2^{l}j$ is equivalent
to $\left(\theta(x)x^{r}\right)^{2^{l}}=\theta(x)x^{r2^{l}}$, since
$\theta(x)$ is an idempotent and $\theta(x)^{2}=\theta(x)$. Consider
$[0,\theta(x)x^{b_{12}},\theta(x)x^{b_{13}},\theta(x)x^{b_{14}}]$
from the $\mathbb{F}_{2}$-rowspace of (\ref{eq:RS15polymatrix}).
Applying the permutation (\ref{eq:RS15bin2}), this row permutes to\\
 \ \ \ $[0,\theta(x)x^{b_{12}},\theta(x)x^{b_{13}},\theta(x)x^{b_{14}}]\rightarrow$\\
 \hspace*{5em} $[\theta(x)x^{2^{l}b_{12}+a_{2}},0,\theta(x)x^{2^{l}b_{14}+a_{4}},\theta(x)x^{2^{l}b_{13}+a_{3}}]$.
\\
 Permutation $\rho$ belong to the code automorphism group if the
following two vector $[\theta(x)x^{2^{l}b_{12}+a_{2}},0,\theta(x)x^{2^{l}b_{14}+a_{4}},\theta(x)x^{2^{l}b_{13}+a_{3}}]$
and $[\theta(x)x^{b_{21}},0,\theta(x)x^{b_{23}},\theta(x)x^{b_{24}}]$
are generated from the same basis vector. This leads to the following
equation 
\begin{equation}
\left\{ \begin{array}{c}
\Delta_{1}=\mathrm{mod}\,(2^{l}b_{12}+a_{2}-b_{\sigma(1),\sigma(2)},n)\\
\Delta_{2}=\mathrm{mod}\,(2^{l}b_{13}+a_{3}-b_{\sigma(1),\sigma(3)},n)\\
\Delta_{3}=\mathrm{mod}\,(2^{l}b_{14}+a_{4}-b_{\sigma(1),\sigma(4)},n)\\
\Delta_{1}=\Delta_{2}=\Delta_{3}
\end{array}\right.
\end{equation}
Let $\Delta_{1}=\Delta_{2}=\Delta$, then $a_{i}=\mathrm{mod}\,(\Delta+b_{\sigma(1),\sigma(i)}-2^{l}b_{1i},n)$.
It can be seen that when $l=0$, we obtain $(a_{2},a_{3},a_{4})=(12,3,9)$.
To verify $\rho$ belong to the code automorphism group, we consider
$[\theta(x)x^{8+r},0,\theta(x)x^{14+r},\theta(x)x^{4+r}]$ from the
$\mathbb{F}_{2}$-rowspace of (\ref{eq:RS15polymatrix}). The permutation
(\ref{eq:RS15bin2}) gives rise to \\
 \ \ \ $[\theta(x)x^{8+r},0,\theta(x)x^{14+r},\theta(x)x^{4+r}]\rightarrow$\\
 \hspace*{10em}$[0,\theta(x)x^{8+r},\theta(x)x^{13+r},\theta(x)x^{2+r}]$\\
 which again lies in the $\mathbb{F}_{2}$-rowspace of (\ref{eq:RS15polymatrix}).
Thus, $\rho$ belongs to the automorphism group of the $(15,12,4)$
RS binary images generated by the matrix in (\ref{eq:parityBinaryImage3parity}).\hfill{}$\Box$

\vspace{0.2cm}
 To find the other permutations of a general $(n,n-3,4)$ RS code,
we propose an effective search permutation algorithm, which is the
generalization of the above example. Note that there exists a set
of constants $\left\{ b_{i,j}:1\leq i,j\leq m,\; i\neq j\right\} $
in $\mathbb{Z}_{n}$, whereby the rows of the following $m\times m$
matrix $\mathbf{M}$: 
\begin{equation}
\mathbf{M}:=\left[\begin{array}{cccc}
0 & \theta(x)x^{b_{12}} & \cdots & \theta(x)x^{b_{1m}}\\
\theta(x)x^{b_{21}} & 0 & \cdots & \theta(x)x^{b_{2m}}\\
\vdots & \vdots & \ddots & \vdots\\
\theta(x)x^{b_{m,1}} & \cdots & \theta(x)x^{b_{m,m-1}} & 0
\end{array}\right]\label{eq:RS2^m_polymatrix}
\end{equation}
lie in the $\mathbb{F}_{2}$-rowspace of (\ref{eq:parityBinaryImage3parity}).
Specifically, (\ref{eq:RS15polymatrix}) in the previous subsection
is but a particular case of (\ref{eq:RS2^m_polymatrix}) with $m=4$
and was reported in \cite{FLim}. The proposed search algorithm is
described below and has a complexity $O(m!n^{2}m)$.

\vspace*{0.5cm}
 \hrule \hspace{1.5cm}\textbf{Heuristic permutation search algorithm}
\hrule \textbf{Input:} Polynomial matrix $\mathbf{M}$ in (\ref{eq:RS2^m_polymatrix}).\\
 \textbf{Output:} The automorphism group $\mathbf{P}$. \\
 1. \textbf{forall} $\sigma\in\Omega_{m},\; a_{2},a_{3}\in\mathbb{Z}_{n}$
and $l\in\mathbb{Z}_{m}$ \\
 2. \hspace*{1em} Compute $\left\{ \begin{array}{c}
\Delta_{1}=\mathrm{mod\:}(2^{l}b_{12}+a_{2}-b_{\sigma(1),\sigma(2)},\: n)\\
\Delta_{2}=\mathrm{mod\:}(2^{l}b_{13}+a_{3}-b_{\sigma(1),\sigma(3)},\: n)
\end{array}\right.$ \\
 3. \hspace*{1em} If $\Delta_{1}\!=\!\Delta_{2}$, then $a_{i}=\mathrm{mod\:}(\Delta_{1}+b_{\sigma(1),\sigma(i)}-2^{l}b_{1i},\: n)$\\
 \hspace*{2.5em} where $3\leq i\leq m$; otherwise \textbf{continue}.
\\
 4. \hspace*{1em} Construct a permutation $\rho=(\sigma,(0,a_{2},\cdots,a_{m}),l)$.\\
 5. \hspace*{1em} If $\mathbf{M}$ is invariant under $\rho$, then
store $\rho$ in $\mathbf{P}$.\\
 6. \textbf{end} \hrule

\vspace{0.5cm}

\textbf{Example 1 cont'd: {[}$(15,12,4)$ RS code{]}} Table \ref{table:AutF15}
presents all the permutations of the $(15,12,4)$ RS binary images
with zeros $\left\{ 1,\alpha,\alpha^{2}\right\} $, obtained by the
proposed algorithm. We report a total of $120$ unique code automorphisms,
$15$ of which map $[i,j]\mapsto[i,j+a]$ for all $a\in\mathbb{Z}_{15}$.
In general, it appears that the permutation $\rho=(\sigma,(a_{1},a_{2},\cdots,a_{m}),0)$,
where $\sigma=(1,n)(2,n-1)\cdots$ and $(a_{1},a_{2},\cdots,a_{m})=(0,(2^{m-1}+1),3,(2^{m-1}+1)+3,6,(2^{m-1}+1)+6,\cdots)+a$
belongs to the automorphism group of the $(n,n-3,4)$ RS binary images
for all $m$ and $a\in\mathbb{Z}_{n}$. \hfill{}$\Box$

\begin{table}[htbf]
\centering \caption{Automorphisms of $(15,12,4)$ RS code with zeros $\left\{ 1,\alpha,\alpha^{2}\right\} $.\label{table:AutF15}}

\vspace{-0.2cm}

\begin{tabular}{|c|c|c|}
\hline 
$\sigma$  & $(a_{1},a_{2},a_{3},a_{4})$  & $l$\tabularnewline
\hline 
$id$  & $(0,0,0,0)+a$  & $0$\tabularnewline
\hline 
$(1,2)(3,4)$  & $(0,12,3,9)+a$  & $0$\tabularnewline
\hline 
$(1,3)(2,4)$  & $(0,3,6,3)+a$  & $0$\tabularnewline
\hline 
$(1,4)(2,3)$  & $(0,9,3,12)+a$  & $0$\tabularnewline
\hline 
$(2,4)$  & $(0,3,9,3)+a$  & $2$\tabularnewline
\hline 
$(4,3,2,1)$  & $(0,0,12,12)+a$  & $2$\tabularnewline
\hline 
$(1,3)$  & $(0,6,0,6)+a$  & $2$\tabularnewline
\hline 
$(1,2,3,4)$  & $(0,12,12,0)+a$  & $2$\tabularnewline
\hline 
\end{tabular}
\end{table}

\textbf{Example 2: {[}$(31,28,4)$ RS code{]}} Consider the $(31,28,4)$
RS binary image with zeros $\left\{ 1,\alpha,\alpha^{2}\right\} $,
where $\alpha$ is a fixed primitive element satisfying $\alpha^{5}=\alpha^{2}+1$,
imaged under the canonical basis. The parity check matrix of this
binary image is invariant under permutation $\rho$, if and only if
the following polynomial matrix in the ring $\mathbb{F}_{2}[x]/(x^{31}-1)$
is invariant under permutation $\rho$: 
\begin{equation}
\left[\begin{array}{ccccc}
0 & \theta(x)x^{6} & \theta(x)x^{12} & \theta(x)x^{15} & \theta(x)x^{24}\\
\theta(x)x^{22} & 0 & \theta(x)x^{5} & \theta(x)x^{11} & \theta(x)x^{14}\\
\theta(x)x^{14} & \theta(x)x^{22} & 0 & \theta(x)x^{5} & \theta(x)x^{11}\\
\theta(x)x & \theta(x)x^{12} & \theta(x)x^{20} & 0 & \theta(x)x^{3}\\
\theta(x)x^{28} & \theta(x)x^{2} & \theta(x)x^{13} & \theta(x)x^{21} & 0
\end{array}\right]\label{eq:RS31polymatrix}
\end{equation}
Applying the proposed search algorithm leads to Table \ref{table:AutF31}.
We identify a total of $124$ unique code automorphisms, $31$ of
which map $[i,j]\mapsto[i,j+a]$ for all $a\in\mathbb{Z}_{31}$.\hfill{}$\Box$

\begin{table}[htbf]
\centering \caption{Automorphisms of $(31,28,4)$ RS code with zeros $\left\{ 1,\alpha,\alpha^{2}\right\} $.\label{table:AutF31}}

\vspace{-0.2cm}

\begin{tabular}{|c|c|c|}
\hline 
$\sigma$  & $(a_{1},a_{2},a_{3},a_{4},a_{5})$  & $l$\tabularnewline
\hline 
$id$  & $(0,0,0,0,0)+a$  & $0$\tabularnewline
\hline 
$(1,2)(4,5)$  & $(0,15,23,29,17)+a$  & $0$\tabularnewline
\hline 
$(1,4)(2,5)$  & $(0,29,9,18,20)+a$  & $0$\tabularnewline
\hline 
$(1,5)(2,4)$  & $(0,17,3,20,6)+a$  & $0$\tabularnewline
\hline 
\end{tabular}
\end{table}

\vspace{0.2cm}
 \textbf{Example 3: {[}General $(n,n-3,4)$ RS codes{]}} The number
of all the automorphism groups found via the proposed algorithm is
listed in Table \ref{table:AutCommonGF}, for several $(n,n-3,4)$
RS codes. We also present the code automorphism orders computed using
MAGMA software. Compared to the existing algorithms, our algorithm
finds significantly more automorphisms than the method in \cite{FLim};
It produces results consistent with those obtained from the MAGMA
computation in \cite{MAGMA}%
\footnote{Note that two groups of the same order do not imply the same.%
}, but can handle fields of larger orders (e.g. $\mathbb{F}_{2^{8}}$)
where MAGMA would face difficulty. \hfill{}$\Box$

\vspace{-0.2cm}

\begin{table}[htbf]
\centering \caption{Automorphism subgroup orders of $(n,n-3,4)$ RS codes over finite
field $\mathbb{F}_{2^{m}}$ obtained by the proposed search algorithm.
(N.A. means not available)}

\vspace{-0.2cm}
 \label{table:AutCommonGF} %
\begin{tabular}{|c|c|c|c|}
\hline 
 & Proposed  & MAGMA \cite{MAGMA}  & Lim \emph{et al.} \cite{FLim} \tabularnewline
\hline 
$\mathbb{F}_{2^{4}}$  & $120$  & $120$  & $18$\tabularnewline
\hline 
$\mathbb{F}_{2^{5}}$  & $124$  & $124$  & N.A\tabularnewline
\hline 
$\mathbb{F}_{2^{6}}$  & $126$  & $126$  & N.A\tabularnewline
\hline 
$\mathbb{F}_{2^{7}}$  & $254$  & $254$  & N.A\tabularnewline
\hline 
$\mathbb{F}_{2^{8}}$  & $510$  & N.A  & N.A\tabularnewline
\hline 
\end{tabular}
\end{table}

\vspace{-0.4cm}

\section{Soft Decoding of $(n,n-3,4)$ RS Codes\label{sec:PermSPA}}

\begin{figure}[htbf]
\centerline{ \includegraphics[scale=0.15]{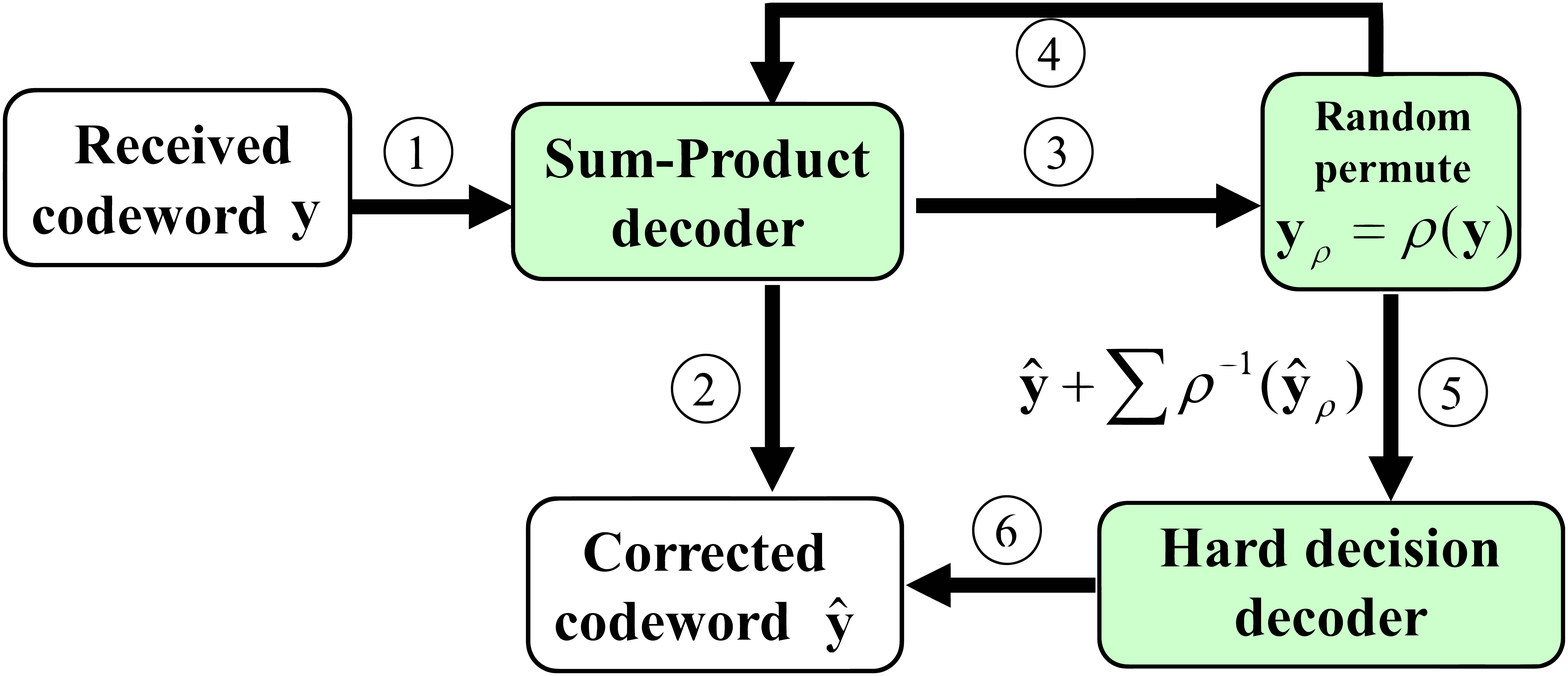} \vspace{-0.5em}
 } \caption{Permutation sum-product algorithm for $(n,n-3,4)$ RS binary images.}

\label{fig:PermSPADecoding} 
\end{figure}

The availability of a list of the permutation groups allows us to
develop more effective soft decoders for triple-parity RS codes. The
proposed new \emph{permutation sum-product algorithm}, illustrated
in Fig. \ref{fig:PermSPADecoding}, takes advantage of the rich permutation
groups obtained in Section \ref{sec:Aut3parity} and the soft decision
output of the classical (binary) sum-product algorithm.

\vspace{0.4cm}
 \hrule \hspace{1cm}\textbf{Permutation sum-product algorithm (PSPA)}
\hrule \textbf{Input:} Observation $\mathbf{y}$ \\
 \textbf{Output:} corrected codeword. \\
 1. Represent $\mathbf{y}$ as a binary image. Decode $\mathbf{y}$
using the conventional sum-product algorithm (SPA).\\
 2. If the SPA produces a valid codeword $\mathbf{c}$, then stop;
otherwise store the log-likelihood $\hat{\mathbf{y}}$ \\
 3. Generate a random permutation $\mathbf{y}$: $\mathbf{y}_{\rho}=\rho(\mathbf{y})$
(e.g. using the proposed search algorithm). \\
 4. Repeat step 1 until all the permutations are exhausted or a maximum
number of trials is reached. \\
 5. Compute the sum: $\mathbf{y}_{sum}=\hat{\mathbf{y}}+\sum\rho^{-1}(\hat{\mathbf{y}}_{\rho})$.\\
 6. Perform hard decision on $\mathbf{y}_{sum}$ and outputs the result.
\hrule

\vspace{0.5cm}
 \textbf{Remark:} It should be noted that our decoding algorithm differs
from that proposed in \cite{Perm_Hehn}. The algorithm in \cite{Perm_Hehn}
is restricted to binary cyclic codes over binary erasure channels.
Our problem here involves the general permutations of bit-level RS
codes over AWGN channels, and the cyclic codes in \cite{Perm_Hehn}
may be viewed as a special case of ours. Our decoder also differs
from that in \cite{FLim}. To achieve its desirable performance, the
decoder in \cite{FLim} must be able to find a ``perfect'' permutation
that gathers all of the bit errors in one symbol, and then decode
this permutation using some \textit{hard-decision} decoder. Not only
is such a perfect permutation very difficult to identify, but the
decoder performance would also degrade very quickly should a perfect
permutation become unavailable. In comparison, our PSPA method obviates
the burden of identifying a perfect permutation, simply takes in random
permutations, and exploits the power of the \textit{soft-decision}
SPA to effectively correct bit errors.\vspace{-0.2cm}

\section{Simulation results}

Simulations are conducted to verify the effectiveness of the proposed
decoding algorithm. The triple-parity $(31,28,4)$ and $(63,60,4)$
RS codes over $\mathbb{F}_{2^{5}}$ and $\mathbb{F}_{2^{6}}$ are
considered. We compare our proposed soft-decision PSPA algorithm with
three other systems: the uncoded system, the RS code with the conventional
hard-decision decoding (HDD) implementing the Berlekamp-Massey algorithm
\cite{Berlekamp}, and the RS code with the soft-decision sum-product
algorithm (SPA) \cite{Pearl}. Since the decoder in \cite{FLim} only
work for double-parity RS codes, it is not included in the comparison.

\begin{figure}[htbf]
\centerline{ \includegraphics[width=3in,height=1.9in]{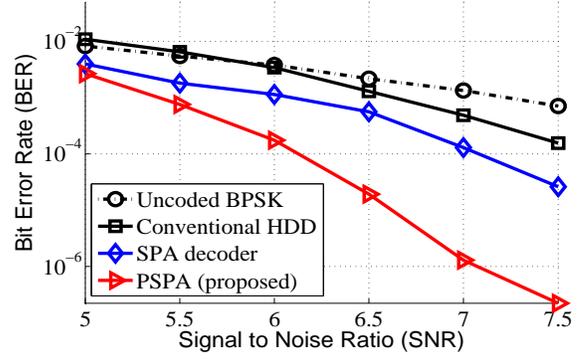}
\vspace{-0.5em}
 } \caption{BER performance for $(31,28,4)$ RS code over AWGN channels.}

\label{fig:BER_PermSPA} 
\end{figure}

\begin{figure}[htbf]
\centerline{ \includegraphics[width=3in,height=1.9in]{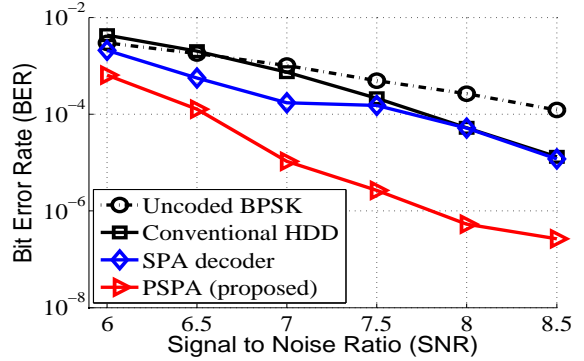}
\vspace{-0.5em}
 } \caption{BER performance for $(63,60,4)$ RS code over AWGN channels.}

\label{fig:BER_PermSPAF64} 
\end{figure}

In the proposed PSPA, the maximum number of decoding iteration is
set to $30$ for the embedded sum-product decoder, and the number
of random permutations (trials) is set to $10$ (i.e. at the most
10 automorphism permutations are chosen randomly from the 4 classes
in Table IV). As shown in Fig. \ref{fig:BER_PermSPA} and Fig. \ref{fig:BER_PermSPAF64},
our PSPA noticeably outperforms the hard-decision decoder and the
soft-decision SPA by more than 1 dB at a BER of $10^{-5}$. This impressive
performance gain is attainable because automorphism permutations lead
to new and different decoding matrices, opening the possibility to
break up some critical short cycles or trapping sets that were previously
preventing the SPA algorithm from converging.

\vspace{-0.3cm}

\section{Conclusion}

We have investigated the bit-level soft decoding of triple-parity
$(n,n-3,4)$ RS codes. We first proposed an effective search algorithm
to identify automorphism groups of a general $(n,n-3,4)$ RS code
with zeros $\left\{ 1,\alpha,\alpha^{2}\right\} $, and next developed
a soft-decoding algorithm that integrates automorphism permutation
and the conventional soft-decision sum product algorithm. Simulations
confirm the effectiveness of the proposed \textit{permutation sum-product
algorithm}, revealing an impressive performance of more than 1 dB
at the BER of $10^{-5}$ than the existing algorithms. Our algorithm
is particularly suitable for short-length triple-parity RS codes,
or concatenated codes in which they are a component code. \vspace{-0.2cm}

\section*{Acknowledgement}

This work is partially funded by the International Design Center (grant
no. IDG31100102 \& IDD11100101) and the Agency for Science, Technology,
and Research, Singapore, Grant No: SERC 0921560129. 

\end{document}